# Studies of a Hollow Cathode Discharge using Mass Spectrometry and Electrostatic Probe Techniques

H. S. Maciel, G. Petraconi, R. S. Pessoa.

*Technological Institute of Aeronautics, Department of Physics – Plasma and Processes Laboratory (LPP), 12228-900, ITA – CTA, São José dos Campos, SP, Brazil.*

Hollow cathode discharges (HCD) are capable of generating dense plasmas and have been used for development of high-rate, low-pressure, high-efficiency processing machines. The geometric feature of a HCD promotes oscillations of hot electrons inside the cathode, thereby enhancing ionization, ion bombardment of inner walls and other subsequent processes. At the same power the hollow cathode exhibits plasma density one to two orders of magnitude higher than that of conventional planar electrodes [1].

The aim of the present studies was to obtain experimental observations about the main features of a dc hollow cathode discharge in order to evaluate its capability of generating compounds in the plasma medium, by reaction between sputtered species from the cathode and radicals from the gas discharge. Especial interest is focused on aluminum nitride (AlN) formation which is very desirable if a deposition of thin film of this material is concerned. The hollow cathode consists of two parallel aluminum plates and the discharge gas was a mixture of argon and nitrogen within the pressure range of 10 to 50 Pa. A Helmholtz coil was used to produce a low intensity and uniform B-field along the axis of the discharge, within the range of $(0 – 30).10^{-3}$ T. The discharge voltage was in the range of $(300 – 900)$ V corresponding to discharge currents in the range $(10 – 800)$ mA. Plasma properties were inferred from the current-voltage characteristics of a single Langmuir probe positioned at the inter-cathode space. For a fixed inter-cathode distance, the electron density and electron temperature were measured for different values of the gas pressure, discharge voltage and magnetic field intensity. Typical values are in the range of $(10^{16} – 10^{17})$ m$^{-3}$ and $(2 – 5)$ eV for the electron density and the electron temperature respectively. Through mass spectrometry technique some species in the plasma gas phase could be monitored for various operating conditions of the discharge. This mass analysis together with the probe measurements gives guidance for optimization of AlN generation in the discharge consequently for deposition of thin films of this material.

## 1. Introduction

High-density plasmas typical of hollow cathodes (HC) discharges are based on efficient avalanche multiplication of electrons known as a hollow cathode effect (HCE) — large increases in current density with reduced separation of the two cathodes. In a dc circuit, this effect is connected with oscillations of electrons between equivalent repelling potentials of sheaths at opposite inner walls in the cathode and consequent effects, for example the ionization of gas, ion bombardment of inner walls with the secondary emission and thermo emission of electrons, etc. Hollow cathode (HC) discharges have too an important characteristic that combines two important processes, i.e. sputtering and excitation/ionization of the sputtered atoms. The HC plasma properties can be enhanced through the use of various geometrics, magnetic field and discharge operating mode modifications [1]. It is known that by controlling the magnetic field geometry close to the discharge [2], the electrical characteristics (discharge current versus discharge voltage) can be modified, for more efficient processing. This feature is particularly useful to increase the discharge current at low running pressures and hence increase the sputter and deposition rates.

In this paper, data obtained from mass spectrometry (MS) and Langmuir electrical probes on a d.c. HC discharge with aluminum cathode in nitrogen and argon/nitrogen are presented and discussed. This investigation has the objective of obtaining preliminary results to evaluate the applicability or the discharge system for deposition of thin films of aluminum nitrite (AlN). Many techniques, such as sputtering, chemical vapor deposition, laser chemical vapor deposition, pulsed laser ablation and molecular beam epitaxy, have been used to fabricate AlN thin films on various substrates. In most of the cases the deposition temperatures are quite high. High temperature deposition has the disadvantage of degradation of the substrate and the AlN thin film during deposition due to thermal damages. Hence, deposition of AlN thin films at low temperatures has become increasingly important and valuable [3]. Sputtering technique is promising under circumstances where temperature deposition low and conformal coatings are needed [4]. As an alternative to the usual magnetron sputtering technique the configuration of planar hollow cathode, which is a much simpler and cheaper system, may have some advantages for the generation of a high density plasma and efficient cathode sputtering. A study of the HC plasma system is therefore desirable to infer its applicability for material processing.

## 2. Experimental

### A. Reactor Configuration

The scheme of the plasma reactor used in this investigation is shown in Fig. 1. The discharge resembles a glow discharge regime extending from the hollow anode (earth potential) up to the flat cathodes made of aluminum. The hollow anode is made of titanium tubing of 7 mm external diameter, covered by a ceramic tube terminated by an orifice of 2.0 mm diameter. The internal diameter of the hollow anode cavity is 4 mm. The orifice constricts the anode discharge to form electrostatic layers which contribute to increase the ionization rate in the anodic region. The vacuum glass chamber of 130 mm internal diameter and 300 mm length was preliminary evacuated using a combination of rotary and diffusion pumps to achieve a residual pressure of $10^{-2}$ Pa. The gas flows were controlled with MKS electronic mass flow meters and the gas pressure was monitored by a MKS capacitance vacuum gauge. The gases were fed through the anode hole at a flow rate within the range of (1-40) sccm. The plane hollow-cathode d.c. discharge was operated with a gas composition of Ar/$N_2$ with pressure within the range of (10-50) Pa. The self-sustained cathode-anode voltage (during operation) was in the range (300-900) V with a discharge current in the range of (10-800) mA. A Helmholtz coil was used to produce a low intensity and uniform B-field along the axis of the discharge, with strength within the range of $(0-30).10^{-3}$ T. The confining magnetic field perpendicular to the cathode plates enhances the pendulum motion of hot electrons between the cathodes thus facilitating the hollow cathode effect, accordingly promoting high rate of cathode sputtering by the impact of plasma ions.

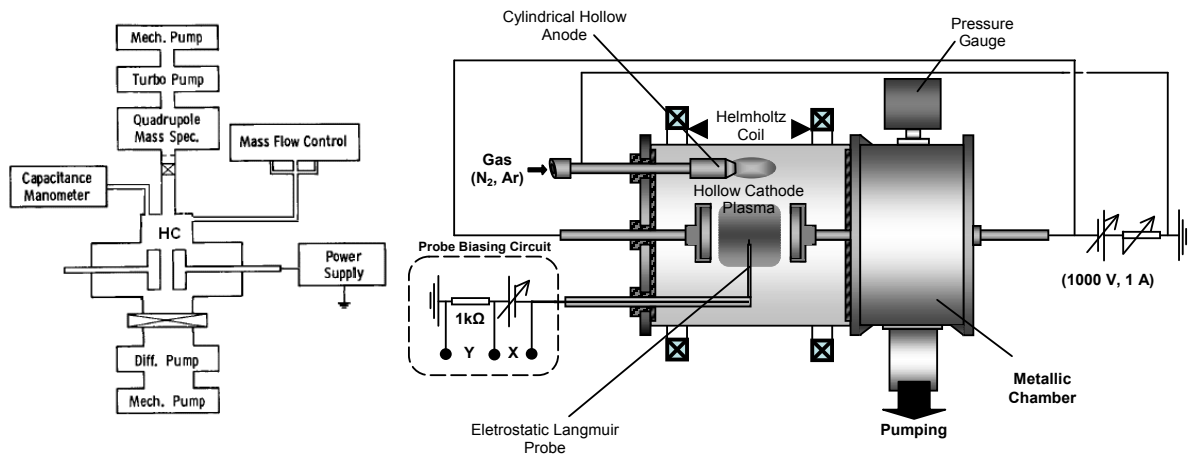

Figure 1. Schematic sketch of the experimental system with cylindrical HC, Residual Gas Analyzer System and a movable single Langmuir probe.

### B. Plasma Diagnostics Techniques

The reactor is equipped with two diagnostic tools, namely:

*1) Langmuir Electrical Probe:* This technique has been utilized to fulfill the need of identifying the macroscopic parameters of the discharge (plasma density, the plasma potential and the electron temperature), which are the real "fingerprints" of a plasma reactor. LEP measurements have been performed by using the configuration of cylindrical single probe. This probe, positioned inside the cathode plasma and at the axis of the discharge, essentially consists of one tungsten wire of 5 mm in length and 0,1 mm in diameter. The unexposed length of the probe was covered with glass and attached to the base of the probe manipulator. The plasma generated between the cathodes is characterized by the presence of impurity metal-ions, gas ions and new compounds due to intensive ion sputtering of the cathode, thus promoting the contamination of the chamber walls and of the electrostatic probes. Prior to *IxV* measurements the probe surface is cleaned by polarizing the probe at the cathode potential during a period short enough for the probe to reach incandescence by ion bombardment heating.

The probe biasing circuit is show in Fig. 1. The points labeled Y input, measuring $I_p$, and X input, measuring $V_p$, correspond to the Y and X channels of the plotter. The circuit is usually a little more complicated, since $V_0$ is not only variable, but must be able to change signs. The voltage can also be swept at a slow rate.

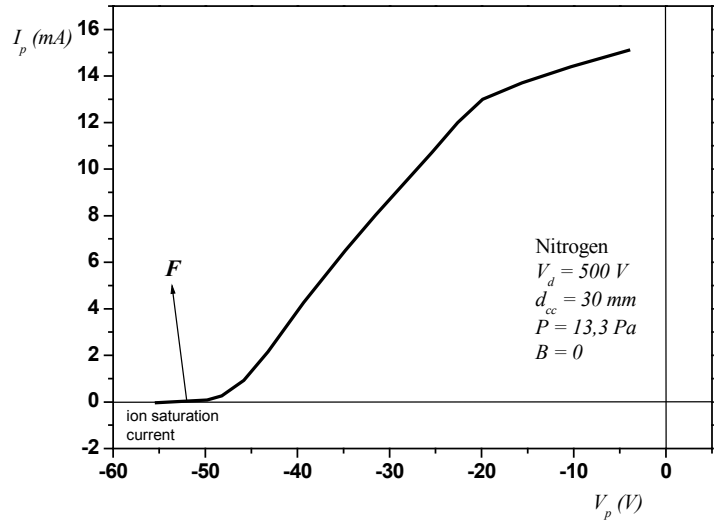

Figure 2. Langmuir probe current ($I_p$) voltage ($V_p$) characteristic. The measurement performed in the HC plasma inter-cathode distance of 30 mm.

Typically output of the probe current ($I_p$) at different applied voltage ($V_p$) is shown in Fig. 2. The analysis of *I-V* curves according to the procedure described in [6] provides the values of electron density ($n_e$) and temperature ($T_e$).

*2) Mass Spectrometer (MS):*

There are nowadays a whole range of techniques available for analyzing glow discharges, for process control and for monitoring deposition rates. With the aim of analyzing the species in the plasma, a mass spectrometer AccuQuad$^{TM}$ with a resolution of 1 amu, was adapted to the vacuum chamber trough a drifting tube for monitoring the mass spectra for different plasma parameters. The species were sampled through a micro orifice located at the mass spectrometer's entrance. The typical operation pressure within the mass spectrometer was $10^{-5}$ Pa. The collected spectra were recorded in the mass range from 1 to 100 amu.

### 3. Results and Discussion

*1) Electron Density and Temperature*

For the determination of the temperature and density of electrons ($n_e$, $T_e$,) a cylindrical probe located in the discharge axis, parallel to the cathode plate as used (see Fig. 3).

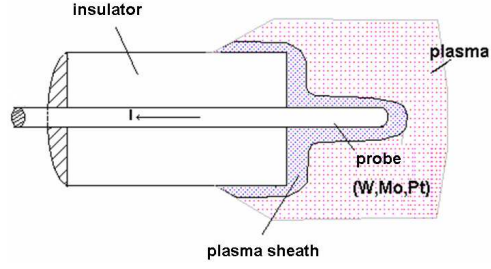

Figure 3. Schematic of a cylindrical probe for collection of the transversal electron flow.

A representative probe I-V characteristic, measured for inter-cathode distance of 30 mm, is given in figure 2. Two important points of the characteristic are $V_f$ (floating potential, $I_p = 0$) and $V_p$ (plasma potential). If the potential of the probe ($V_s$) is negative with respect to the plasma potential, then the number density of the electrons that can reach the probe, assuming Boltzmann electron energy distribution, is

$$n_e = n_{eo} \exp(\frac{eV}{kT}) \qquad (1)$$

In terms of the electron current:

$$I_s = eSn_{eo}\sqrt{\frac{kT_{e\perp}}{2\pi m_e}} \exp(\frac{eV}{kT_e}) \qquad (2)$$

where $S$ is the probe area and $I_{eo} = eSn_{oe}\sqrt{\frac{kT_e}{2\pi m_e}}$ is the electron saturation current collected by the probe when its potential equals the plasma potential. In the logarithmic form the equation (2) becomes

$$\ln(I_s) = \ln(I_{eo}) + \frac{eV}{kT_e} \qquad (3)$$

The value of $V_p$, which can be easily read from the plot of $lnI_p$ versus $V_p$ (Fig. 4) is about -39 V with respect to earth. The floating potential $V_f$ measured at the point $F$ (Fig. 2) is about -52 V. This means that the net floating potential measured in relation to the plasma ($V_f$-$V_p$) is about -13 V.

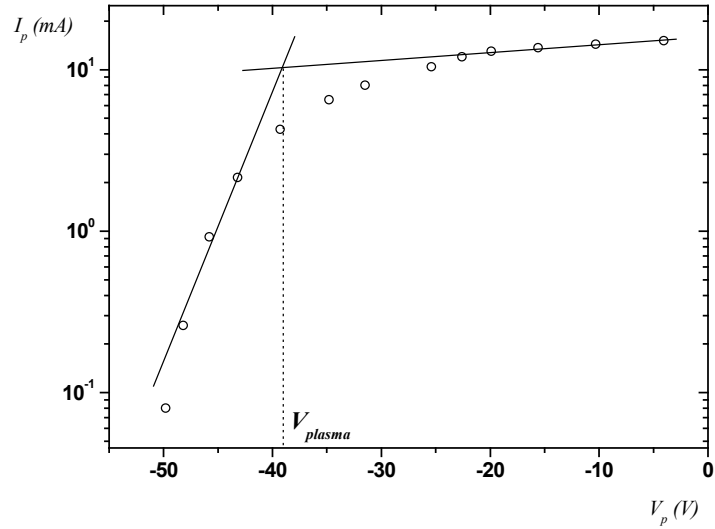

Figure 4. The *I-V* characteristic (from Fig. 2) plotted in semilog coordinate system.

An electron temperature of 2,8 eV is determined from the slope of the *I-V* characteristic plotted in the semilog coordinate system (Fig. 4).

The electron saturation current which corresponds to the point of plasma potential, $V_p$ is given by.

$$I_{oe} = n_o eS \sqrt{\frac{kT_e}{2\pi m_e}} \qquad (4)$$

where *S* is the probe surface area, $m_e$ and *e* are the electron mass and electric charge.

The density of electrons $n_o \approx n_{eo}$ is determined from the expression (4) by using the value of $T_e$. In this case we obtain $n_0 = 7{,}0 \times 10^{16}$ m$^{-3}$.

*2) Dependence of the plasma parameters on the inter-cathode distance*

The temperature and density of the plasma electrons nearby the ionic sheath determine the positive ion flux towards a surface. Therefore, the temperature and density of the electrons of the plasma in the vicinity of the deposition substrate are relevant plasma parameters in determining the structure and surface properties of the deposited film. In this section results concerning the effect of the d.c. discharge power and longitudinal magnetic field, on the temperature and density of the plasma electrons in the nitrogen plasma are reported.

It is well known that the product (*Pd*), of the inter-cathode distance (*d*) by the pressure (*P*), is an important parameter to describe the behavior of the HC discharge. Usually, the electron-atom inelastic collision rates are increased by the decrease of the inter-cathode distance with a large effect on the plasma density and electron temperature. The effect of the gas pressure on the discharge properties is expected since the increase of the collisionality by increasing the pressure tends to enhance the hollow cathode effects being possible to reach an optimized reduced inter-cathode distance (*Pd*). Probe measurements were performed at gas pressure of 13,3 Pa and inter-cathode distance within the range of (20 – 65)mm for constant values of the discharge voltage, maintaining fixed the nitrogen flow rate. Fig. 5-a shows the dependence of plasma density on *Pd* for three values of the discharge voltage. In general we observe that the curves have a descendent behaviour with the density decreasing being more pronounced in the branch of low values of *Pd*. In this branch, the density increases sharply with a decreasing inter-cathode distance (for fixed pressure) because the high energy electrons, which are emitted by the cathode, oscillate between repelling potential of the sheaths at opposite cathode walls. These electrons enhance ionization in the negative glow and in the sheaths. Thus, due to the effect of electrostatic electron confinement, at low values of *Pd*, the plasma assumes high values of electron and ion density thus reverting in higher sputtering rates of the cathode material. The discharge current, fig. 5-b, exhibits similar behavior.

(a) (b)

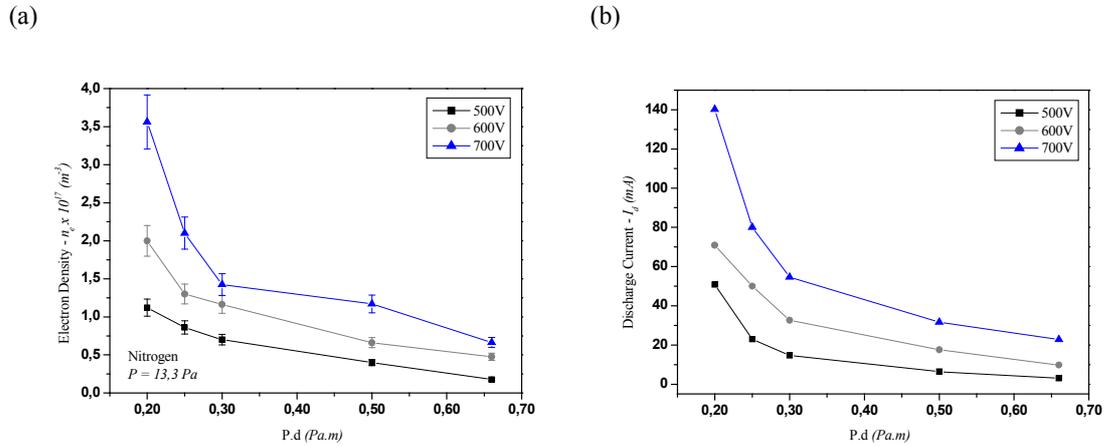

Figure 5. Dependence of the electron density and discharge current on the product *Pd* for three values of the discharge voltage.

The application of a weak magnetic field perpendicularly to the cathode surface facilitates the ignition of the discharge and enhances - depending on the operating pressure and the inter-cathode distance - the hollow cathode effect by promoting an increase of the residence time of electrons in the cathodic cavity and by reducing the lateral diffusion. Consequently the discharge current and electron density increase with the application of a magnetic field as can be observed in the figures 6 a-b.

(a) (b)

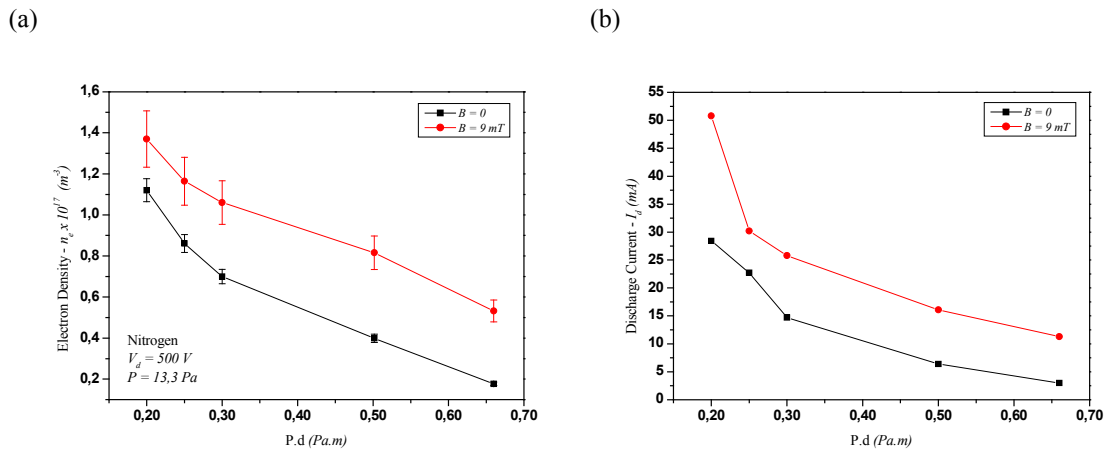

Figure 6. Effect of the magnetic field on the electron density and discharge current for constant values of the discharge voltage and gas pressure.

The nature of the gas affects also the discharge properties as illustrated in fig.7. The hollow cathode effect is more pronounced for oxygen and less pronounced for argon.

Figure 8 presents the dependence of the electron temperature on the inter-cathode distance at three values of the discharge power for nitrogen discharge. By increasing the inter-cathode distance, the electron temperature decreases, as expected: at low inter-cathode distances, the electrons lose little energy in inelastic collisions with the atoms/molecules and their mean value of the kinetic energy (electron temperature) is high.

As the distance *d* increases, more collisions occur, the electron kinetic energy loss through electron inelastic collisions with neutral atoms/molecules increases thus the electron temperature decreases. Also, because of their high electron affinity, the $N_2$ atoms capture low energy electrons from plasma and form negative ions with the effects of increasing of temperature of the plasma electrons for low *Pd* values.

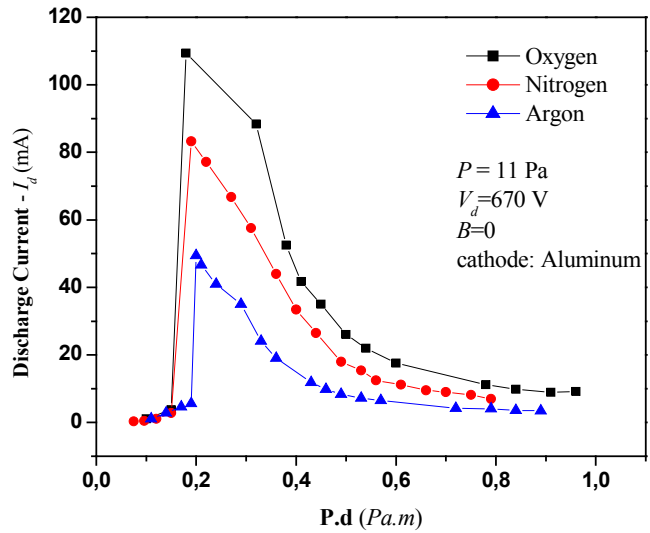

Figure 7. Effect of the gas type on hollow cathode discharge current at constant value of the discharge voltage.

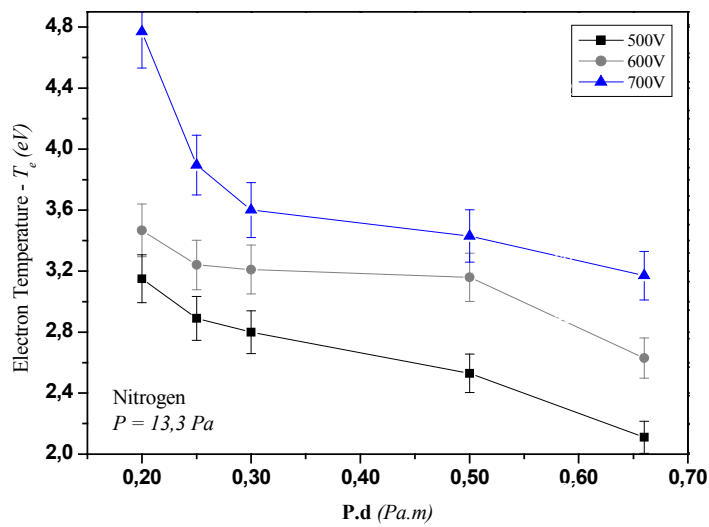

Figure 8. Variations of the electron temperature with *Pd*, for different values of the HC discharge voltage.

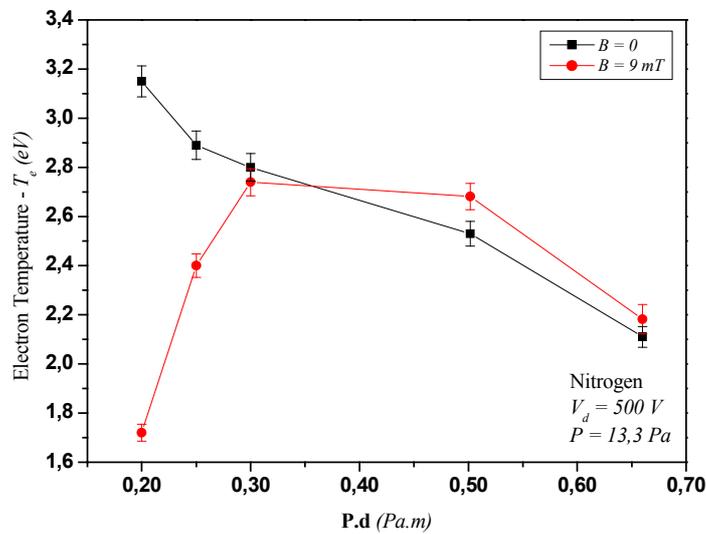

Figure 9. Effect of the magnetic field on the electron temperature at a constant value of the discharge voltage.

A substantial decreasing of the measured electron temperature was observed when a magnetic field was applied, as shown in Fig. 9. This effect is not totally related to the plasma electrons temperature decrease only but probably due to the effect of magnetic field on the electron flux reaching the probe. Especially for low values of $Pd$ the hot oscillating electrons (pendulum effect) have a more directional motion – perpendicular to the cathode surfaces – and the magnetic field favors this orientation. On the other hand it is known that the application of a magnetic field on a glow discharge has the equivalent effect of an increase of the gas pressure thereby a decrease of the effective electron temperature.

*3) Plasma composition analysis by mass spectrometry*

A large number of variables affect the balance between competing chemical and physical processes in discharge plasma. Appropriate reliable diagnostic tools are therefore necessary to provide an understanding of the plasmas and the actual phenomena taking place, which in turn will enable the control of the process. Mass spectrometry has been shown to be one very useful tool for diagnostics of cold plasmas.

In this investigation, our main interest is to search for evidences of aluminum nitride (AlN) compounds formation in the plasma medium of a HC discharge made of aluminum cathode and running with nitrogen gas. A typical mass spectrum of sampled gaseous components formed in the discharge is presented in figure 10. The partial pressure is plotted as function of the mass with partial pressure on a logarithmic scale to make possible the comparison of the peak intensities within a range of four orders of magnitude. In the atomic mass unit (amu) range of 1 to 100 we observe many peaks and among them we can distinguish: hydrogen molecule at 2, water that leads to the primary peaks at 17 and 18 due to the species $O^+$, $HO^+$, and $H_2O^+$, nitrogen molecule at mass 28 which also causes the peaks at 14 assigned to atomic nitrogen $N^+$ and doubly ionized $N_2^{++}$. Molecular oxygen is shown a peak at 32. Carbon dioxide appears with a peak at 44 and also the peaks for $CO_2^{++}$ and $C^+$ at 22 and 12 respectively. There are other peaks caused by fragments of these species and contaminants. The line at m/z = 41, assigned to the aluminum nitride molecule is clearly seen. The appearance of $AlN^+$ peak corroborates our hypothesis that this compound is present in the plasma medium and may contribute, besides of the aluminum atom (at peak 27) and nitrogen radicals, to the mass flux reaching a substrate where AlN film is grown.

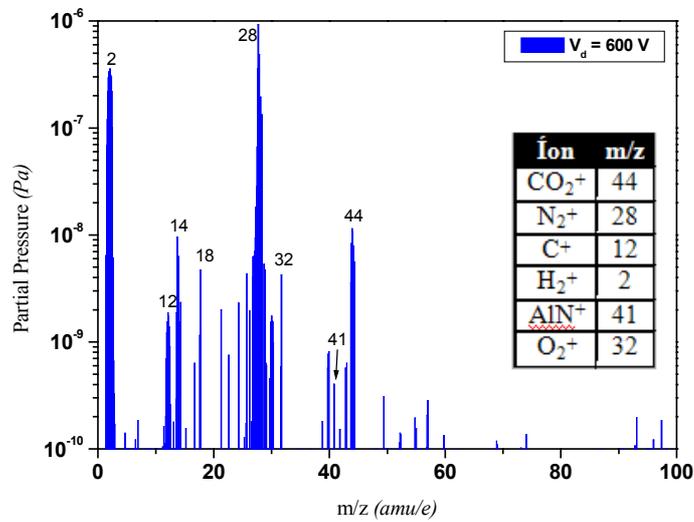

Figure 10. Mass spectrum of the discharge in $N_2$ for pressure of 13,3 Pa, subtracted from background spectrum.

The AlN$^+$ line can be monitored as function of the operating parameters of the discharge. In the Figs. 11 and 12 the relative percentage of AlN$^+$ and Al$^+$ peak height were plotted as function of the inter-cathode distance, for two values of pressure ($P$). For discharge processes where the operation pressure is lower (fig.11), the curve indicates that the generation of aluminum nitride molecules reaches a maximum for a certain inter-cathode distance. This maximum is less pronounced at lower pressure (fig. 12).

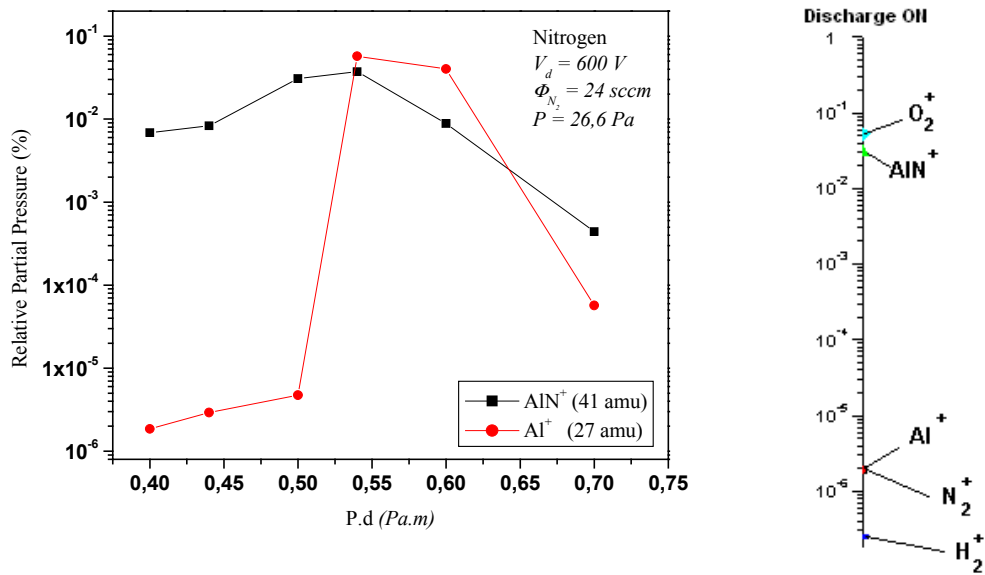

Figure 11. Variation of the Aluminum (27 amu) and AlN$^+$ (41 amu) peaks for the hollow cathode discharge operating with nitrogen at pressure of 26.6Pa.

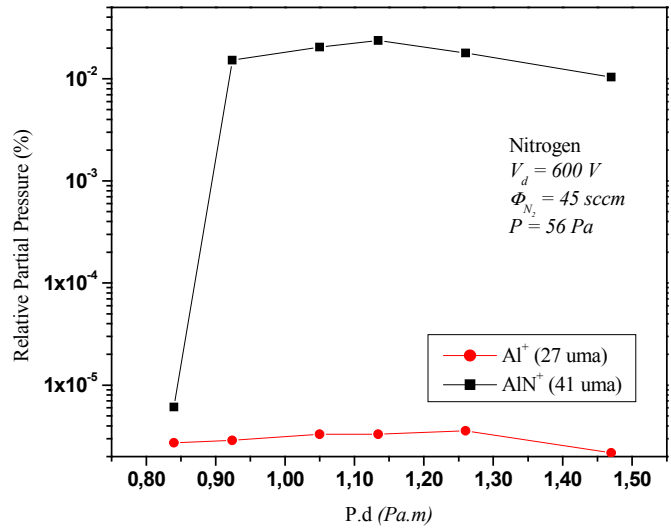

Figure 12. Variation of the Aluminum (27 amu) and AlN⁺ (41 amu) peaks with the discharge operating at a nitrogen pressure of 56 Pa.

By varying the percentage of nitrogen in the argon/nitrogen mixture we observe, from figure 13, that the AlN peak height can vary within three orders of magnitude. This figure indicates also that, as far as the deposition rate of AlN thin films is concerned, the percentage o nitrogen to be used should be in the range of 50% to 100%

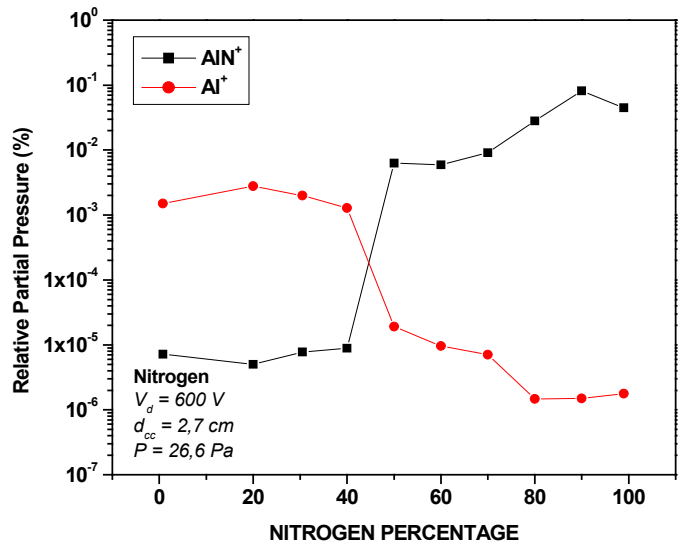

Figure 13. The variation of the aluminum nitrite and atomic aluminum signals as a function of the percentage of nitrogen in the $N_2$/Ar gas mixture.

## 4. Conclusions

Investigations were made on a hollow aluminum cathode discharge operating with nitrogen and argon/nitrogen. A cylindrical Langmuir probe was used to measure the density and temperature of the electrons of hollow cathode nitrogen plasma. The dependence of these plasma parameters on the inter-cathode distance, discharge voltage and magnetic field was experimentally determined. The plasma has electron density of the order of $10^{16}$ m$^{-3}$ and electron temperature in the range of (2 – 5) eV. Through mass spectrometry technique various compounds generated in the discharge were identified. An important result of this work was the demonstration of the presence of aluminum nitride (AlN) molecule in the plasma. Due to ion bombardment and an enhanced cathode temperature, aluminum atoms are sputtered from the cathode and react with nitrogen radicals to form AlN$^+$ in the plasma medium. The results obtained by monitoring the AlN peak of the mass spectrum combined with Langmuir probe measurements can indicate the optimal operating conditions of the hollow cathode system for enhanced generation of AlN molecules in the plasma and hence for growing thin films of aluminum nitride.

**References**

[1] D. Zhechev, V.I. Zhemenik, S. Tileva, G.V. Mishinsky, N. Pyrvanova, "A hollow cathode discharge modification as a source of sputtered atoms and their ions", Nuclear Inst. and Methods in Physics Research, B 204, (2003) 387.
[2] Kadlec, S. and Musil, J. J., Vat. Sci. Technol., 1996, A13 (2), 389.
[3] H. Chenga, Y. Sunb, J.X. Zhangb, Y.B. Zhangc, S. Yuanb, P. Hinga, "AlN films deposited under various nitrogen concentrations by RF reactive sputtering", J. of Crystal Growth 254 (2003) 47.
[4] R. Bathe, R.D. Vispute, Dan Habersat, R.P. Sharma, T. Venkatesan, C.J. Scozzie, M. Ervin, B.R. Geil, A.J. Lelis, S.J. Dikshit, R. Bhattacharya, Thin Solid Films 398 (2001) 575.
[5] N. V. Garrilov, G. A. Hesyats, G. V. Radkovski, V. V. Bersenev, "Development of Technological Sources of gas ions on the basis of Hollow-Cathode Glow Discharge", Surface and Coatings Technology, 96 (1997) 81 – 88.
[6] F. F. Chen, in Plasma Diagnostic Techniques, edited by R. H. Huddlestone and S. L. Leonard (Academic, New York, (1965), Chap. 4.